# Structure and Bonding in Amorphous Red Phosphorus

Yuxing Zhou,[a] Stephen R. Elliott,[b] and Volker L. Deringer[a,*]

[a] Department of Chemistry, Inorganic Chemistry Laboratory, University of Oxford, Oxford OX1 3QR, United Kingdom

[b] Department of Chemistry, Physical and Theoretical Chemistry Laboratory, University of Oxford, OX1 3QZ, United Kingdom

*E-mail: volker.deringer@chem.ox.ac.uk

**Amorphous red phosphorus (a-P) is one of the remaining puzzling cases in the structural chemistry of the elements. Here, we elucidate the structure, stability, and chemical bonding in a-P from first principles, combining machine-learning and density-functional theory (DFT) methods. We show that a-P structures exist with a range of energies slightly higher than those of phosphorus nanorods, to which they are closely related, and that the stability of a-P is linked to the degree of structural relaxation and medium-range order. We thus complete the stability range of phosphorus allotropes [Angew. Chem. Int. Ed. 2014, *53*, 11629] by now including the previously poorly understood amorphous phase, and we quantify the covalent and van der Waals interactions in all main phases of phosphorus. We also study the electronic densities of states, including those of hydrogenated a-P. Beyond the present study, our structural models are expected to enable wider-ranging first-principles investigations – for example, of a-P-based battery materials.**

## Introduction

Phosphorus is one of the structurally most diverse elements, and its various allotropes continue to attract widespread research interest in chemistry.[1] White phosphorus is the thermodynamic standard state and consists of tetrahedral $P_4$ molecules.[2] Black phosphorus shows a layered structure with buckled six-membered rings held together by van der Waals (vdW) dispersion interactions,[3] and can be exfoliated to form monolayer "phosphorene", which is beginning to be used for multiple advanced technologies.[4] Hittorf's violet[5] and Ruck's fibrous phosphorus[6] both contain cage-like fragments, with five-membered rings as the principal building unit, and characteristic "P8" and "P9" cages being found in both modifications. These fragments connect to form perpendicular (parallel) tubular structures in violet (fibrous) phosphorus, respectively. Similar cages and tubular networks are found in recently synthesised phosphorus nanorods[7] and nanowires.[8]

In addition to the crystalline allotropes, there is a widely-known non-crystalline form, namely, red amorphous phosphorus (a-P).[9] An emerging application for this material is in batteries,[10] based on its ability to form Li–P and Na–P phases that lead to high theoretical capacities of a-P-based anodes, as long as the conductivity and the volumetric change during cycling can be controlled.[11] In terms of structural chemistry, various models have been proposed for a-P, including two-dimensional structures with layered motifs, similar to those in black phosphorus;[12] tubular networks containing cage-like motifs, *e.g.*, P8 and P9 fragments;[9b] or a structural model primarily composed of P3 rings and P4 tetrahedra that form extended chains.[13] (We here write "P4" rather than "$P_4$", for consistency of notation.) However, whilst the crystalline structures can be accurately characterised in advanced diffraction and imaging experiments,[6, 14] a large part of the difficulty in studying a-P is in determining its structure in the first place. Early work using neutron diffraction suggested the existence of P8 or P9 motifs, inferred from a similarity to Hittorf-type fragments,[9b] whereas empirical potential structure refinement based on X-ray diffraction data also implied the presence of P4 tetrahedra.[13] In addition, Raman spectroscopy studies suggested the existence of both buckled six-membered rings[15] and cage-like motifs,[16] indicating a rather complex atomic structure of a-P, whose details may well depend on the synthesis conditions.

We have recently shown that machine-learning (ML) methods which are "trained" on quantum-mechanical reference data can lead to an unprecedented level of quality in describing a-P on the atomic scale.[17] Specifically, we created an a-P structural model (containing 1,984 atoms) by simulated slow cooling from a disordered melt.[18] The resultant structure primarily contains cluster fragments of five-membered rings, in line with the long-established Baudler rules[19] and with the foundational theoretical work on gas-phase clusters by Böcker and Häser.[20] The validity of the structural model was shown by comparison to the available experimental evidence: the simulated first sharp diffraction peak (FSDP) in the structure factor of our model matches previous experimental results very well, and so does its evolution in compression and decompression simulations – see Ref. [18] and references therein.

And yet, just like there are open questions about many crystalline phosphorus allotropes,[1,21] there remain fundamental chemical questions about the amorphous form, a-P. For example, how does the degree of local structural ordering determine the energetic (meta-) stability? What is the chemical-bonding nature, which might be expected to include both strong covalent and weaker dispersion interactions? What is the role of coordination defects? And where does a-P fall within the previously introduced first-principles stability range of the crystalline phosphorus allotropes?[21]

In the present work, we address precisely those questions by using a suite of state-of-the-art computational chemistry techniques: ML-driven molecular-dynamics (MD) simulations, first-principles DFT including many-body-dispersion (MBD) corrections, as well as analyses of electronic structure and orbital interactions. We introduce a series of structural models for a-P and also for its hydrogenated form (a-P:H) – large enough to represent the complex structures, yet small enough to enable full first-principles investigations. We derive new insight into the role of local structural order in the energetic stability of a-P, the nature of defects, and the balance between covalent and vdW dispersion interactions in phosphorus allotropes.

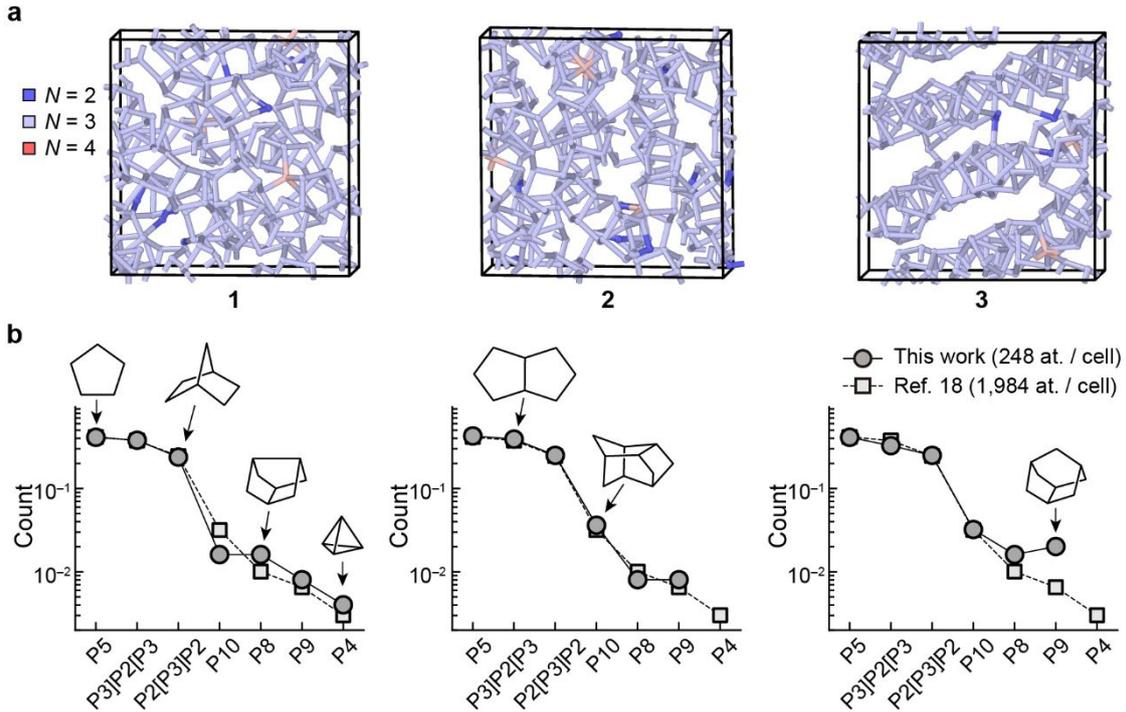

**Figure 1.** Structural models of amorphous phosphorus. (**a**) Structures of a-P as obtained from ML-driven simulations and subsequent DFT structural optimisation. We generated three independent structural models containing 248 atoms each, using the same simulation protocol in each case. The structures are labelled **1** to **3**, with an increasing degree of structural order. Colour-coding indicates the coordination numbers, *viz.* $N = 2$ (blue), $N = 3$ (pale blue), and $N = 4$ (pink), determined using a 2.4 Å cut-off. (**b**) Relative abundance (per-atom count) of cluster fragments in the a-P models (circles), as compared to the result of a similar but larger-scale 1,984-atom simulation in Ref. [18] (squares). Solid and dashed lines are guides to the eye. Sketches of selected fragments are shown, following Ref. [20].

## Results and Discussion

We carried out melt-quench MD simulations to create structural models of a-P.[18] Our simulations start from a metastable liquid, rather than the fluid (consisting of P4 molecules) which is stable at low pressure – this is done on purpose, because the large structural diversity in the metastable liquid allows the simulation to rapidly explore relevant local structural fragments. The protocol followed our earlier work,[18] but the new simulations were now carried out in smaller simulation cells, with the aim to enable subsequent DFT studies, and in three parallel runs. Having three small-scale models of a-P allows us to take advantage of the statistical nature of the process: with the same protocol but different starting configurations, these structures over millions of simulation steps evolved into rather distinct disordered networks, which we label **1** to **3** (Figure 1).

Our a-P models, after DFT-based structural optimisation, contain mostly three-fold coordinated atoms, in line with standard valence rules; there are only ≈ 1% of over-coordinated ($N = 4$) and ≈ 2% of under-coordinated atoms ($N = 2$). There is a different degree of structural ordering in the three a-P models (Figure 1a): **1** shows a more disordered, random network, whereas the arrangement of fragments in **3** follows patterns resembling those in violet and fibrous phosphorus (*viz.* tubular networks). Similar to the previously simulated large-scale a-P model,[18] the three models created in this work contain abundant five-membered rings, and more complex cluster fragments that are made up of those (*e.g.*, the P8 cage found in violet and fibrous P consists of four fused five-membered rings). We use the same notation as in Ref. [20] to label those fragments.

The distribution of cluster fragments (Figure 1b), together with a similarity in short-range structural features (Figure S1c–d), confirms that our new small-scale models are overall consistent with the previously validated, larger-scale model. A comparatively higher count of P8 and P9 cages in **3** implies a greater similarity with the closely related crystalline forms, violet and fibrous phosphorus. Such cage-like fragments have also been used as building blocks of predicted phosphorus allotropes with more complex local environments, *via* a random structure search.[22] No P4 units are observed in **2** and **3**, implying pure network, rather than molecular, structures. The overall agreement between the present structures and the one from Ref. [18] underlines that purpose-tailored simulations are possible with ML potentials: enabling ultra-large-scale simulations, but also the use of smaller-scale models (*e.g.*, **1**–**3** in this work) which are accessible to subsequent first-principles DFT analyses of energetics and chemical bonding.

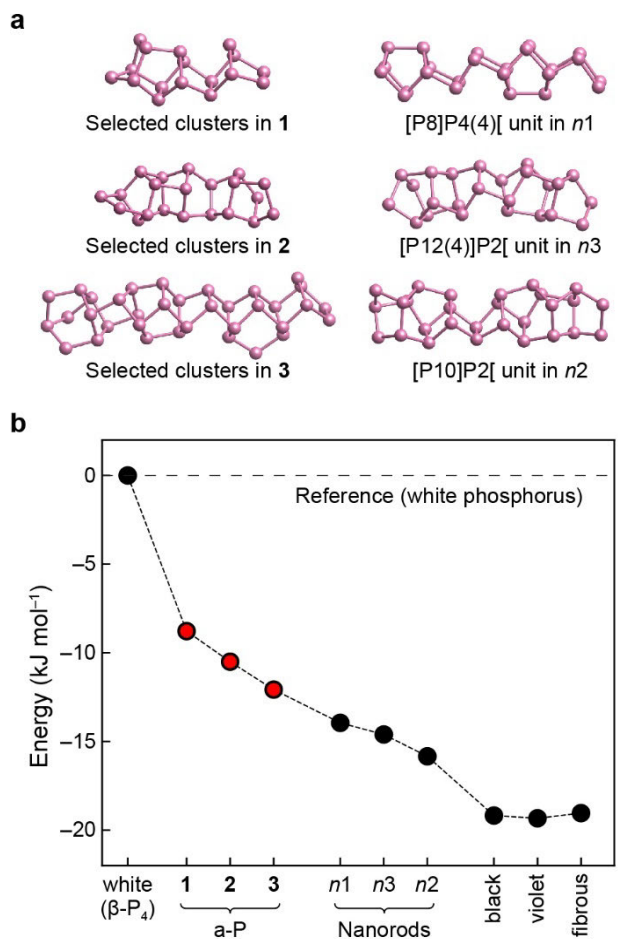

**Figure 2.** The energetic stability range of phosphorus modifications, including the well-known white, black, violet, and fibrous forms, as well as structural models of Pfitzner's nanorods (*n*1 to *n*3) based on Refs. [7] and [21], and the new a-P structural models **1** to **3** generated in the present work. (**a**) Selected, local structural fragments in our a-P models compared to the building units of nanorods. (**b**) Computed energies, given relative to the standard state, *viz.* white phosphorus, and based on DFT computations (HSE06+MBD).

To compare the energetic stability of different phosphorus phases, including our a-P structures (**1** to **3**) and relevant crystalline forms, we computed the energies of fully DFT-optimised structures for those modifications (Figure 2). Corrections for vdW interactions are needed to accurately describe longer-range interactions in phosphorus modifications:[17, 23] we here used the "D3" method[24] for structural optimisations, and the many-body dispersion (MBD) method[25] for subsequent single-point computations. Both methods describe the exfoliation of black phosphorus remarkably well,[26] and MBD also captures the unit-cell volume of black phosphorus in almost quantitative agreement with experiment, and the exfoliation energy as compared to higher-level quantum-chemistry methods,[17] *viz.* quantum Monte Carlo[27] and coupled-cluster theory.[28]

The computed energies provide a full stability range of phosphorus allotropes. Among the phases studied, white phosphorus is the least energetically stable (Figure 2). Violet, fibrous, and black phosphorus are all predicted to have close energies, with differences less than 0.3 kJ mol$^{-1}$, although the structure of black phosphorus differs from that of the other two. In fact, such energetic near-degeneracy is not only seen with HSE06+MBD, but was reported using various DFT methods.[21, 29] Following Ref. [21], we also generated three models of P nanorods (referred to as *n*1, *n*2, and *n*3), by removing the copper and iodine atoms from $(CuI)_8P_{12}$, $(CuI)_3P_{12}$, and $(CuI)_2P_{14}$, respectively, and then relaxing the structures using DFT. These models consist of more complex tubular chains and are energetically less stable than the "textbook" crystalline allotropes. Our a-P models (**1** to **3**), energetically, sit in between white P and the other crystalline forms, and with increased structural ordering (Figures 1a and S2), the energies of these three models decrease. Table 1 shows that the computed mass densities of different allotropes agree well with experimental data, and the computed energy ranking of the crystalline forms is consistent with previous studies[21, 23] – our combined ML- and DFT-based simulation approach has now allowed us to extend this ranking to the previously poorly understood amorphous form.

Bachhuber *et al.* have previously studied dispersion interactions in phosphorus allotropes, using pairwise dispersion-correction methods available at the time.[32] With advanced MBD corrections, we also observe that the vdW contributions to the overall energy vary across the different phosphorus modifications (Table 1): white phosphorus shows the smallest absolute vdW contribution (lowering the total energy by about 14 kJ mol$^{-1}$), whereas black phosphorus has the largest value ($\approx$ 20 kJ mol$^{-1}$). All structures studied that contain cage-like motifs have a similar level of vdW contributions, of about 16–17 kJ mol$^{-1}$ in stabilisation relative to the pure, uncorrected HSE06-DFT energy. These results can be understood from the different interspaces of building fragments in molecular and network solids: in white phosphorus, the P4 tetrahedra are well-separated from each other, and the distance between building units is longer than in other modifications, resulting in smaller dispersion contributions. This result is intuitive in the sense that vdW forces approximately decay with the sixth power of the distance, but it is still notable that the "molecular" solid, P4, is less strongly vdW bonded than any of the black, violet, fibrous, or amorphous forms.

**Table 1.** Properties of the a-P models **1** to **3** and comparison to crystalline phases. The standard phase, white phosphorus, was set as the reference. The relative energies of other structures are given with respect to white phosphorus.

|  | $\Delta E$ (kJ mol$^{-1}$) |  | $\rho$ (g cm$^{-3}$) |  |
|---|---|---|---|---|
|  | Total | vdW (MBD) contribution | DFT | Expt. |
| White (β-P$_4$) | ± 0 (reference) | –13.7 | 1.89 | 1.98[2] |
| **1** | -8.8 | –16.4 | 2.25 |  |
| **2** | –10.5 | –16.5 | 2.27 | 2.14 to 2.34[13, 30] |
| **3** | –12.1 | –16.5 | 2.29 |  |
| *n*1 | –14.0 | –15.9 | 2.18 |  |
| *n*2 | –15.8 | –16.8 | 2.29 |  |
| *n*3 | –14.6 | –16.4 | 2.23 |  |
| violet | –19.3 | –17.1 | 2.32 | 2.36[5a] |
| fibrous | –19.0 | –17.2 | 2.31 | 2.37[6] |
| black | –19.2 | –19.9 | 2.64 | 2.69[31] |

We next created structural models for hydrogenated a-P (referred to as "a-P:H"), aiming to better understand atomic coordination defects (*i.e.*, over- and under-coordinated atoms) and how they might affect the electronic properties of a-P. Comparable hydrogenated phases of amorphous silicon (known as "a-Si:H") have long been studied: the material can be generated under synthesis conditions (*e.g.*, plasma decomposition of silane) in which atomic defects are passivated by hydrogen atoms,[33] leading to varying concentrations of hydrogen and defects in the material.[34] Experiments have shed light on the activation and functionalisation of white phosphorus[35] to form different P-containing species (*e.g.*, organophosphorus compounds);[36] under specific synthesis conditions, P4 tetrahedra break and give rise to unsaturated phosphorus atoms, which readily form new bonds. We here considered hydrogenated models for a-P, following a similar idea as in a-Si:H, as unsaturated ($N = 2$) phosphorus atoms may also be expected to be passivated during synthesis (*e.g.*, when a-P is formed by thermal decomposition of $PH_3$).[37]

Figure 3a illustrates the hydrogenation process, starting from pristine a-P (models **1** to **3**), and leading to three hydrogenated structures which we label **1H** to **3H**. We generated these structures as follows: (1) all fourfold connected atoms (≈ 1% of the total atoms in the model) were removed; (2) the valences of all the resulting twofold connected atoms were saturated by adding one additional hydrogen atom to each, forming P–H bonds perpendicular to the two P–P bonds and on the side with more open space; (3) the modified structures were further relaxed using dispersion-corrected DFT (PBE+D3; Methods section). This procedure leaves the connectivity of different cluster fragments intact, whilst ensuring that all P atoms are three-fold connected in the resulting hydrogenated models.

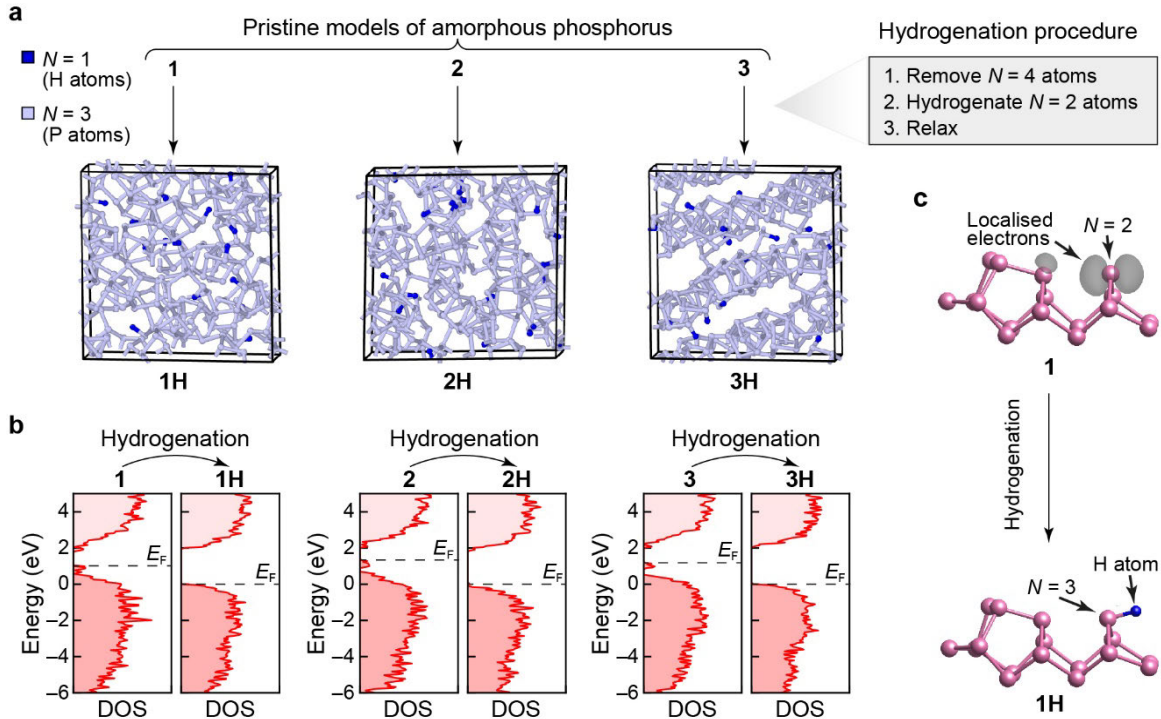

**Figure 3.** Hydrogen in a-P and its effect on the electronic structure. (**a**) Structural models for a-P:H, **1H** to **3H**. To generate these structural models, four-fold-connected atoms were removed and resulting two-fold-connected atoms saturated with hydrogen atoms, followed by relaxation at the PBE+D3 level. Consequently, these structures are fully connected according to standard valence rules. (**b**) Computed electronic density of states (DOS) plots for pristine and hydrogenated a-P, using HSE06. Dashed lines indicate the DFT-computed Fermi level, $E_F$, here taken to be located at the top of the valence band in the hydrogenated models. The energy scales (*y*-axis) for the pristine models were each shifted such that the zero energy represents $E_F$ of the corresponding hydrogenated model. (**c**) Visualisation of the band-resolved charge density for the highest occupied band (grey isosurfaces) of a defect state in model **1** (upper panel) and its corresponding state (lower panel) in model **1H** after hydrogenation. The coordination defect ($N = 2$) in **1** and the same atom (now fully connected; $N = 3$) in **1H**, together with the added hydrogen atom, are marked by arrows. Except for the marked atoms, all other atoms are fully connected ($N = 3$) in the structural fragments shown. Different isovalues are used (*viz.* 0.006 in **1**; and 0.001 in **1H**), to only highlight the charge distribution around the marked atoms. Despite the lower isovalue used for **1H**, no localised electrons were found on visual inspection, corroborating the removal of defect states upon hydrogenation.

The electronic densities of states (DOS) for the pristine and hydrogenated a-P models (Figure 3b) were computed using hybrid DFT. All three pristine models show mid-gap defect states between the valence-band maximum and the conduction-band minimum. The band-resolved charge densities for the highest occupied bands (Figure 3c) show that these mid-gap states are mostly caused by non-bonded electrons around the twofold-connected atoms and by the defect states at the fourfold-connected atoms in pristine a-P models. Upon hydrogenation, all defect states disappeared, increasing the band gaps in all three hydrogenated models compared to their pristine counterparts. The **3H** model has a slightly larger predicted gap (2.06 eV) than **1H** (2.01 eV) and **2H** (1.86 eV), suggesting that structural ordering might play a role in opening up the electronic band gap in a-P. The computed band gaps for our a-P models agree well with various, previously reported experimental results for amorphous red phosphorus (1.42 to 2.07 eV).[38]

Expanding upon prior studies of crystalline allotropes,[5b, 39] we also calculated the electronic DOS, at the hybrid-DFT level, for all crystalline phases discussed in this work (Figure S3) – allowing for side-by-side comparison with our a-P models. A wide range of band gaps was found for the various phosphorus modifications: white P has the largest gap (3.80 eV) among the crystalline allotropes, consistent with its molecular nature, whereas only a small gap exists in black P (0.23 eV), in agreement with experimental data ($\approx$ 0.3 eV).[15, 40] By contrast, violet and fibrous phosphorus have moderate gap sizes (2.26 and 2.30 eV, respectively), close to that of **3H**, which shows a relatively more ordered network than the other two hydrogenated models. The band gap sizes of the nanorod structures are in the same range as for other forms of P containing cage-like motifs (*viz.* 2.01–2.26 eV, Figure S3).

The Crystal Orbital Hamilton Population (COHP) technique allows one to quantify the bond strength and understand the bonding features, based on orbital interactions from the hybrid DFT calculations. We calculated –COHP curves for all P–P bonds in the crystalline and amorphous modifications mentioned above; negative COHP values stand for stabilisation (bonding). The integrated –COHP (referred to as –ICOHP) of a given bond, up to the Fermi level, $E_\text{F}$, has been used to quantify the strength of chemical bonds in similar systems.[41] Figure 4a shows the relative bond strength, normalised to the shortest bond in crystalline (violet) phosphorus, for all crystalline modifications and up to 2.7 Å. An empirical measure, introduced by Pauling in the 1940s, can be used to describe this relation between bond length and strength:[42]

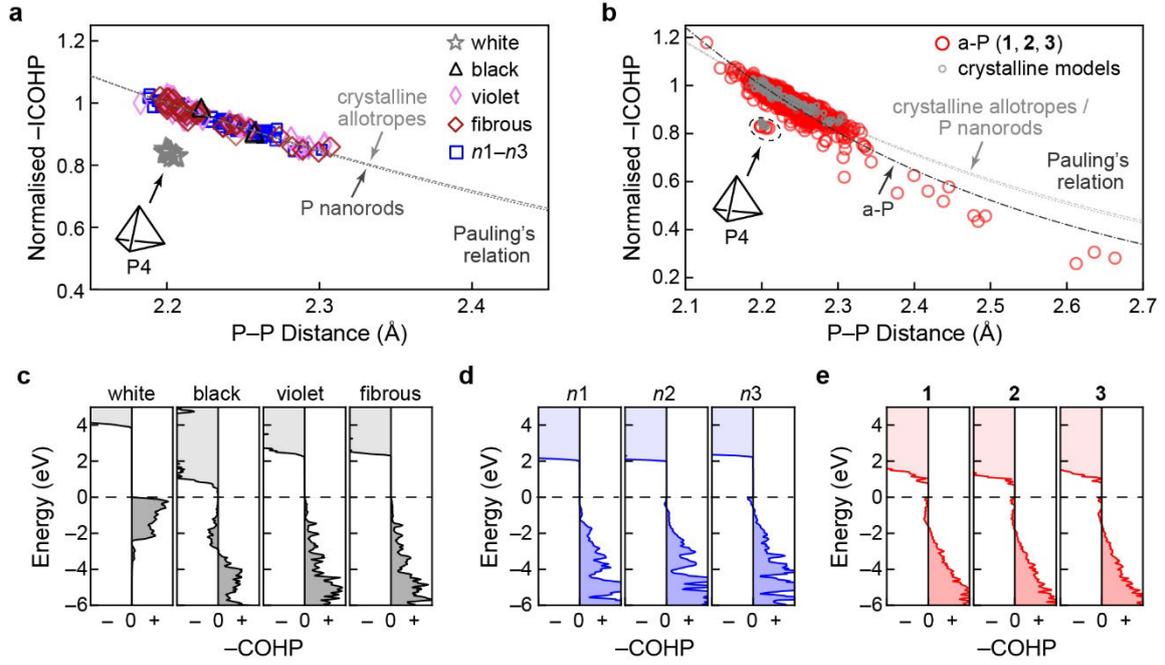

**Figure 4.** Chemical bonding in crystalline and amorphous phosphorus. Bond-length–bond-strength correlations in the different modifications are obtained from integrated crystal orbital Hamilton population (ICOHP) analysis of: (**a**) crystalline modifications, including the nanorod models *n*1–*n*3; and (**b**) a-P models. In these plots, ICOHP values are all normalised using a factor of 6.07 eV, such that the shortest bond (2.18 Å) in crystalline violet phosphorus has a value of 1.0. White phosphorus stands apart from the other crystalline models, whereas the rest of the crystalline structures and nanorods all follow a similar trend. Curves representing Pauling's relation are shown by dashed lines (see text). We note that the bonds in the single P4 unit in the a-P model **1** are located in the same region as those for crystalline white P in panel (a); in both cases, these data points have not been included in the fits. (**c**–**e**) Energy-resolved –COHP plots, emphasising the close similarity between a-P, nanorods, and violet and fibrous phosphorus. The positive and negative signs on the abscissae indicate the bonding and anti-bonding regions, respectively. The dashed horizontal line indicates the Fermi level, $E_\mathrm{F}$.

$$D1 - D = c \, \log_{10} n$$

in which $D1$ is the single P–P bond length based on tabulated covalent radii (2.22 Å),[42] and $D$ is the length of different P–P bonds in the system. $n$ is the normalised –ICOHP value, and $c$ is a fitting coefficient. During fitting, data for white phosphorus were not included, since its bonding nature, involving P4 molecules, clearly deviates from that of the other crystalline modifications. Bonds longer than 2.7 Å were also ignored, as Pauling's formula is only expected to describe well the region of relatively strong covalency.

We found that the fitted Pauling relation for black, violet, and fibrous phosphorus almost overlaps with the one fitted for P nanorods, suggesting similar bonding in all these crystalline modifications. The same analysis was performed for all P–P bonds in the three pristine a-P models (Figure 4b). The –ICOHP data for a-P scatter in a wide range due to more complex bonding environments; the bonds in the single P4 unit in model **1** are located away from the others, in the same region as found for crystalline white phosphorus. Fitting Pauling's relation based on the rest of the bonds suggests slightly different bonding in a-P than in the crystalline phases: the bond strength diminishes more quickly with bond length in a-P, perhaps indicating slightly lower chemical stability of the extended network.

Energy-resolved –COHP plots provide more detailed "fingerprints" of bonding, as shown in Figure 4c. Despite stable bonds being formed in white phosphorus (within the P4 tetrahedra), evident from stabilising areas below $E_F$ (–COHP > 0), the overall bond strength as measured by –ICOHP is lower than in the other forms. By contrast, a small *anti*bonding interaction below $E_F$ is observed for black phosphorus (–COHP < 0). Such interactions were reported for P–P bonds in structurally related compounds, *viz.* (Li-intercalated) phosphorene[43] and BaP$_4$Te$_2$,[44] and in various chalcogenides, such as the iso-valent-electronic GeTe[45] and related Ge–Sb–Te alloys.[46] Hence, antibonding areas below $E_F$ do not necessarily suggest poor stability, especially as there are no antibonding interactions directly at $E_F$, and the –ICOHP value is large. Violet and fibrous phosphorus show only bonding areas below $E_F$, and so does *n*1 (with P8 building units similar to those in the violet and fibrous forms); more complex cage motifs (*e.g.*, P10 and P12) might result in marginal antibonding regions at the valence-band edge in *n*2 and

*n*3. Similar interactions are seen for a-P: there are small antibonding regions not only for the defect states in the gap, but also below the top of valence band. The former is largely attributed to mid-gap defect states; the latter may be due to complex cage motifs (such as in *n*2 and *n*3) that are less "ideal" than the P8 / P9 building units in the violet and fibrous forms.

## Conclusions

We have created structural models of amorphous red phosphorus by combining machine-learning-driven molecular dynamics and first-principles computations. Our a-P models are energetically intermediate between white phosphorus and the other crystalline modifications. We find that the details of their energetic stability depend on the degree of structural ordering: the less stable amorphous model **1** is rather disordered and resembles a random network, whereas the more stable model **3** is structurally similar to crystalline tubular phases (*e.g.*, violet and fibrous phosphorus). Our work completes the first-principles investigation of the stability range of phosphorus modifications,[21] having added the challenging case of a-P to the picture. We also created models of hydrogenated models a-P, thereby revealing the impact of defect states on the electronic properties of pristine a-P. Our chemical-bonding analyses quantified the relation between bond length and strength, indicating slightly weaker covalent bonding in a-P than in its crystalline counterparts. Whilst the present work has focused on the fundamental structural chemistry of a-P, the structural models provided herein are amenable to further DFT investigations: as one example, we envision first-principles computational studies of Na-ion insertion in chemically more complex a-P-based battery materials.

## Methods

**GAP-driven melt-quench simulations.** A general-purpose ML potential, introduced in Ref. [17], was used to generate structural models of a-P. For this potential, the Gaussian Approximation Potential (GAP) framework[47] was used, together with the Smooth Overlap of Atomic Positions (SOAP)[48] structural descriptor. The reference database, from which the potential had "learned", contains configurations of various crystalline, nanostrucured, and liquid phases, and random structures generated via a GAP-driven random structure-searching protocol.[49] The reference data (energies, forces, and stresses) had been computed at the PBE+MBD level.

We performed GAP-driven MD simulations in the NPT ensemble, using LAMMPS,[50] with a Nosé–Hoover thermostat[51] controlling the temperature and a barostat[52] controlling the external pressure. Melt-quench simulations were used to generate small-scale a-P models, starting from three metastable liquid configurations taken from the reference database of the original potential paper.[17] These models each contain 248 atoms with mass densities of 2.5, 2.5, and 2.4 g cm$^{-3}$, respectively; their structures represent different stages of the liquid-to-liquid phase transition.[53] Both P molecules and covalently connected networks were observed in the former two models, and the third almost fully resembles the network liquid without any P tetrahedra. The models were rapidly de-compressed from $\approx$ 1 to 0 GPa and annealed at 1,500 K over 10 ps, and then quenched to 1,200 K at a rate of $10^{13}$ K s$^{-1}$. After that, the structural models were slowly cooled down from 1,200 K to 100 K at a rate of $10^{11}$ K s$^{-1}$, thereby forming amorphous networks (Figure S1a). During quenching, the number of fully three-coordinated atoms steadily increased (Figure S1b). We have previously shown that this process allows the simulation to explore the relevant configurational space, leading to a structural model consistent with previous experimental data (including structure factors at ambient and high pressure; see Ref. [18] and references therein). The a-P models were further relaxed successively using GAP and DFT (PBE+D3). A force tolerance of 0.01 eV Å$^{-1}$ was used as the stopping criterion in GAP-based relaxations.

**Structural models of hydrogenated a-P.** To obtain structural models of hydrogenated a-P (a-P:H) without coordination defects, over-coordinated ($N = 4$) P atoms in the pristine models were removed, leading to the formation of additional two-coordinated P atoms. An exception to this exists when an $N = 4$ atom is bonded to an $N = 2$ atom. Instead of removing the $N = 4$ atom, which leads to an $N = 1$ atom with a dangling P–P bond, the $N = 2$ atom was removed. After that, one hydrogen atom was added to each of the under-coordinated ($N = 2$) P atoms, placed perpendicularly to the two chemical bonds of that P atom, on the side with more open space (*i.e.* with the lowest sum of distances to all neighbouring P atoms in the local environment), forming a short P–H bond. The initial bond length was set to 1.4 Å, a typical value found in phosphorus hydrides.[54] The resultant structures were then computationally optimised using PBE+D3.

**DFT computations.** Structural relaxations and single-point calculations were performed using the Vienna Ab initio Simulation Package (VASP)[55] with projector augmented-wave (PAW)[56] pseudopotentials. The PBE+D3 method[24, 57] was used for structural relaxations, whereas the HSE06 hybrid functional[58] with many-body dispersion (MBD) corrections[25] was used in the subsequent static computations, based on pre-converged PBE wave functions. The plane-wave energy cut-off was 500 eV, the energy tolerance for SCF convergence was $10^{-7}$ eV per cell, and the force tolerance for structural relaxation was 0.01 eV Å$^{-1}$. Gaussian smearing with a width of 0.05 eV was used to determine partial occupancies during SCF cycles. In structural relaxation, all lattice parameters and atomic coordinates were optimised. A *k*-point grid with a largest allowed spacing of 0.1 Å$^{-1}$ along each reciprocal lattice vector was used for structural relaxation for crystalline modifications, and a spacing of 0.15 Å$^{-1}$ (*i.e.*, a lower *k*-point density) was used for the following HSE06 computations of the electronic structure and bonding, to reduce computational cost. We verified that changing between both settings did not change the predicted band gap of black phosphorus (0.23 eV) within the quoted accuracy. For the a-P and a-P:H models, all data reported are from Γ-point-only computations.

**Chemical-bonding.** The Local-Orbital Basis Suite Towards Electronic-Structure Reconstruction (LOBSTER) code[59] was used to project the self-consistent wave functions onto an auxiliary basis of localised, atom-centered orbitals (3s and 3p on each P atom), enabling crystal orbital Hamilton population (COHP) analysis.[60]

**Visualisation.** The structural models in Figure 1a and 3a, as well as the fragments shown in Figure 2a, were visualised using OVITO;[61] the visualisations of local motifs and band-resolved charge density in Figure 3c were created using VESTA.[62]

## Acknowledgements

Y.Z. acknowledges a China Scholarship Council-University of Oxford scholarship. This work was performed using resources provided by the Cambridge Service for Data Driven Discovery (CSD3) operated by the University of Cambridge Research Computing Service (www.csd3.cam.ac.uk), provided by Dell EMC and Intel using Tier-2 funding from the Engineering and Physical Sciences Research Council (capital grant EP/P020259/1), and DiRAC funding from the Science and Technology Facilities Council (www.dirac.ac.uk). The authors would also like to acknowledge the use of the University of Oxford Advanced Research Computing (ARC) facility in carrying out this work (http://dx.doi.org/10.5281/zenodo.22558).

## Conflict of Interest

The authors declare no conflict of interest.

## Data Availability Statement

The data supporting the present study will be made publicly available through the Zenodo repository upon journal publication.

## References

[1] A. Pfitzner, *Angew. Chem. Int. Ed.* **2006**, *45*, 699–700.

[2] A. Simon, H. Borrmann, J. Horakh, *Chem. Ber.* **2006**, *130*, 1235–1240.

[3] X. Ling, H. Wang, S. Huang, F. Xia, M. S. Dresselhaus, *Proc. Natl. Acad. Sci. U. S. A.* **2015**, *112*, 4523–4530.

[4] a) H. Liu, A. T. Neal, Z. Zhu, Z. Luo, X. Xu, D. Tomanek, P. D. Ye, *ACS Nano* **2014**, *8*, 4033–4041; b) L. Li, Y. Yu, G. J. Ye, Q. Ge, X. Ou, H. Wu, D. Feng, X. H. Chen, Y. Zhang, *Nat. Nanotechnol.* **2014**, *9*, 372–377.

[5]   a) H. Thurn, H. Krebs, *Angew. Chem. Int. Ed.* **1966**, *5*, 1047–1048; b) L. Zhang, H. Huang, B. Zhang, M. Gu, D. Zhao, X. Zhao, L. Li, J. Zhou, K. Wu, Y. Cheng, J. Zhang, *Angew. Chem. Int. Ed.* **2020**, *59*, 1074–1080.

[6]   a) M. Ruck, D. Hoppe, B. Wahl, P. Simon, Y. Wang, G. Seifert, *Angew. Chem. Int. Ed.* **2005**, *44*, 7616–7619; b) N. Eckstein, A. Hohmann, R. Weihrich, T. Nilges, P. Schmidt, *Z. Anorg. Allg. Chem.* **2013**, *639*, 2741–2743.

[7]   A. Pfitzner, M. F. Brau, J. Zweck, G. Brunklaus, H. Eckert, *Angew. Chem. Int. Ed.* **2004**, *43*, 4228–4231.

[8]   J. B. Smith, D. Hagaman, D. DiGuiseppi, R. Schweitzer-Stenner, H. F. Ji, *Angew. Chem. Int. Ed.* **2016**, *55*, 11829–11833.

[9]   a) W. L. Roth, W. T. De, A. J. Smith, *J. Am. Chem. Soc.* **1947**, *69*, 2881–2885; b) S. R. Elliott, J. C. Dore, E. Marseglia, *J. phys., Colloq.* **1985**, *46*, C8.

[10]  a) Y. Kim, Y. Park, A. Choi, N. S. Choi, J. Kim, J. Lee, J. H. Ryu, S. M. Oh, K. T. Lee, *Adv. Mater.* **2013**, *25*, 3045–3049; b) J. Qian, X. Wu, Y. Cao, X. Ai, H. Yang, *Angew. Chem. Int. Ed.* **2013**, *52*, 4633–4636; c) I. Capone, J. Aspinall, E. Darnbrough, Y. Zhao, T.-U. Wi, H.-W. Lee, M. Pasta, *Matter* **2020**, *3*, 2012–2028.

[11]  W. Li, Z. Yang, M. Li, Y. Jiang, X. Wei, X. Zhong, L. Gu, Y. Yu, *Nano Lett.* **2016**, *16*, 1546–1553.

[12]  a) J. S. Lannin, B. V. Shanabrook, *Solid State Commun.* **1978**, *28*, 497–500; b) J. S. Lannin, B. V. Shanabrook, F. Gompf, *J. Non-Cryst. Solids* **1982**, *49*, 209–219.

[13]  J. M. Zaug, A. K. Soper, S. M. Clark, *Nat. Mater.* **2008**, *7*, 890–899.

[14]  W. Lu, X. Ma, Z. Fei, J. Zhou, Z. Zhang, C. Jin, Z. Zhang, *Appl. Phys. Lett.* **2015**, *107*, 021906.

[15]  Z. Yang, J. Hao, S. Yuan, S. Lin, H. M. Yau, J. Dai, S. P. Lau, *Adv. Mater.* **2015**, *27*, 3748–3754.

[16]  a) D. J. Olego, J. A. Baumann, M. A. Kuck, R. Schachter, C. G. Michel, P. M. Raccah, *Solid State Commun.* **1984**, *52*, 311–314; b) G. Fasol, M. Cardona, W. Hönle, H. G. von Schnering, *Solid State Commun.* **1984**, *52*, 307–310.

[17]  V. L. Deringer, M. A. Caro, G. Csányi, *Nat. Commun.* **2020**, *11*, 5461.

[18]  Y. Zhou, W. Kirkpatrick, V. L. Deringer, *Adv. Mater.* **2022**, *34*, 2107515.

[19]  M. Baudler, *Angew. Chem. Int. Ed.* **1982**, *21*, 492–512.

[20]  S. Böcker, M. Häser, *Z. Anorg. Allg. Chem.* **1995**, *621*, 258–286.

[21] F. Bachhuber, J. von Appen, R. Dronskowski, P. Schmidt, T. Nilges, A. Pfitzner, R. Weihrich, *Angew. Chem. Int. Ed.* **2014**, *53*, 11629–11633.

[22] V. L. Deringer, C. J. Pickard, D. M. Proserpio, *Angew. Chem. Int. Ed.* **2020**, *59*, 15880–15885.

[23] M. Aykol, J. W. Doak, C. Wolverton, *Phys. Rev. B* **2017**, *95*, 214115.

[24] a) S. Grimme, J. Antony, S. Ehrlich, H. Krieg, *J. Chem. Phys.* **2010**, *132*, 154104; b) S. Grimme, S. Ehrlich, L. Goerigk, *J. Comput. Chem.* **2011**, *32*, 1456–1465.

[25] a) A. Tkatchenko, R. A. DiStasio, Jr., R. Car, M. Scheffler, *Phys. Rev. Lett.* **2012**, *108*, 236402; b) A. Ambrosetti, A. M. Reilly, R. A. DiStasio, Jr., A. Tkatchenko, *J. Chem. Phys.* **2014**, *140*, 18A508.

[26] G. Sansone, A. J. Karttunen, D. Usvyat, M. Schutz, J. G. Brandenburg, L. Maschio, *Chem. Commun.* **2018**, *54*, 9793–9796.

[27] L. Shulenburger, A. D. Baczewski, Z. Zhu, J. Guan, D. Tománek, *Nano Lett.* **2015**, *15*, 8170–8175.

[28] M. Schutz, L. Maschio, A. J. Karttunen, D. Usvyat, *J. Phys. Chem. Lett.* **2017**, *8*, 1290–1294.

[29] A. Impellizzeri, A. A. Vorfolomeeva, N. V. Surovtsev, A. V. Okotrub, C. P. Ewels, D. V. Rybkovskiy, *Phys. Chem. Chem. Phys.* **2021**, *23*, 16611−16622.

[30] V. V. Brazhkin, A. J. Zerr, *J. Mater. Sci.* **1992**, *27*, 2677–2681.

[31] A. Brown, S. Rundqvist, *Acta Crystallogr.* **1965**, *19*, 684–685.

[32] F. Bachhuber, J. von Appen, R. Dronskowski, P. Schmidt, T. Nilges, A. Pfitzner, R. Weihrich, *Z. Kristallogr. Cryst. Mater.* **2015**, *230*, 107–115.

[33] a) V. L. Deringer, N. Bernstein, A. P. Bartok, M. J. Cliffe, R. N. Kerber, L. E. Marbella, C. P. Grey, S. R. Elliott, G. Csányi, *J. Phys. Chem. Lett.* **2018**, *9*, 2879–2885; b) J. Robertson, *J. Appl. Phys.* **2000**, *87*, 2608–2617; c) A. H. M. Smets, W. M. M. Kessels, M. C. M. van de Sanden, *Appl. Phys. Lett.* **2003**, *82*, 1547–1549.

[34] a) G. D. Cody, T. Tiedje, B. Abeles, B. Brooks, Y. Goldstein, *Phys. Rev. Lett.* **1981**, *47*, 1480–1483; b) H. Shanks, C. J. Fang, L. Ley, M. Cardona, F. J. Demond, S. Kalbitzer, *Phys. Status Solidi B* **1980**, *100*, 43–56; c) A. A. Langford, M. L. Fleet, B. P. Nelson, W. A. Lanford, N. Maley, *Phys. Rev. B* **1992**, *45*, 13367–13377.

[35] C. M. Hoidn, D. J. Scott, R. Wolf, *Chem. Eur. J.* **2021**, *27*, 1886–1902.

[36] a) U. Lennert, P. B. Arockiam, V. Streitferdt, D. J. Scott, C. Rodl, R. M. Gschwind, R. Wolf, *Nat. Catal.* **2019**, *2*, 1101–1106; b) D. J. Scott, J. Cammarata, M. Schimpf, R. Wolf, *Nat. Chem.* **2021**, *13*, 458–464.

[37] a) P. Abraham, A. Bekkaoui, V. Souliére, J. Bouix, Y. Monteil, *J. Cryst. Growth* **1991**, *107*, 26–31; b) M. Ceppatelli, D. Scelta, M. Serrano-Ruiz, K. Dziubek, G. Garbarino, J. Jacobs, M. Mezouar, R. Bini, M. Peruzzini, *Nat. Commun.* **2020**, *11*, 6125.

[38] a) K. Kawashima, H. Hosono, Y. Abe, S. Fujitsu, H. Yanagida, *J. Non-Cryst. Solids* **1987**, *95–96*, 741–747; b) D. Xia, Z. Shen, G. Huang, W. Wang, J. C. Yu, P. K. Wong, *Environ. Sci. Technol.* **2015**, *49*, 6264–6273; c) S. A. Ansari, M. S. Ansari, M. H. Cho, *Phys. Chem. Chem. Phys.* **2016**, *18*, 3921–3928; d) Y. Xuan, H. Quan, Z. Shen, C. Zhang, X. Yang, L. L. Lou, S. Liu, K. Yu, *Chem. Eur. J.* **2020**, *26*, 2285–2292.

[39] a) Y. L. Lu, S. Dong, J. Li, Y. Wu, L. Wang, H. Zhao, *Phys. Chem. Chem. Phys.* **2020**, *22*, 13713–13720; b) P.-L. Gong, D.-Y. Liu, K.-S. Yang, Z.-J. Xiang, X.-H. Chen, Z. Zeng, S.-Q. Shen, L.-J. Zou, *Phys. Rev. B* **2016**, *93*, 195434; c) J. Guan, W. Song, L. Yang, D. Tománek, *Phys. Rev. B* **2016**, *94*, 045414.

[40] J. Kim, S. S. Baik, S. H. Ryu, Y. Sohn, S. Park, B. G. Park, J. Denlinger, Y. Yi, H. J. Choi, K. S. Kim, *Science* **2015**, *349*, 723–726.

[41] a) V. L. Deringer, R. P. Stoffel, M. Wuttig, R. Dronskowski, *Chem. Sci.* **2015**, *6*, 5255–5262; b) A. L. Görne, R. Dronskowski, *Carbon* **2019**, *148*, 151–158.

[42] L. Pauling, *J. Am. Chem. Soc.* **1947**, *69*, 542–553.

[43] G. Xu, Y. Liu, J. Hong, D. Fang, *2D Mater.* **2020**, *7*, 025028.

[44] S. Jorgens, D. Johrendt, A. Mewis, *Chem. Eur. J.* **2003**, *9*, 2405–2410.

[45] U. V. Waghmare, N. A. Spaldin, H. C. Kandpal, R. Seshadri, *Phys. Rev. B* **2003**, *67*, 125111.

[46] M. Wuttig, D. Lüsebrink, D. Wamwangi, W. Wełnic, M. Gilleßen, R. Dronskowski, *Nat. Mater.* **2007**, *6*, 122–128.

[47] a) A. P. Bartók, M. C. Payne, R. Kondor, G. Csányi, *Phys. Rev. Lett.* **2010**, *104*, 136403; b) V. L. Deringer, A. P. Bartók, N. Bernstein, D. M. Wilkins, M. Ceriotti, G. Csányi, *Chem. Rev.* **2021**, *121*, 10073–10141.

[48] A. P. Bartók, R. Kondor, G. Csányi, *Phys. Rev. B* **2013**, *87*, 184115.

[49] V. L. Deringer, D. M. Proserpio, G. Csányi, C. J. Pickard, *Faraday Discuss.* **2018**, *211*, 45–59.

[50] A. P. Thompson, H. M. Aktulga, R. Berger, D. S. Bolintineanu, W. M. Brown, P. S. Crozier, P. J. in 't Veld, A. Kohlmeyer, S. G. Moore, T. D. Nguyen, R. Shan, M. J. Stevens, J. Tranchida, C. Trott, S. J. Plimpton, *Comput. Phys. Commun.* **2022**, *271*, 108171.

[51] a) W. G. Hoover, *Phys. Rev. A* **1985**, *31*, 1695–1697; b) S. Nosé, *Mol. Phys.* **1984**, *52*, 255–268.

[52] W. Shinoda, M. Shiga, M. Mikami, *Phys. Rev. B* **2004**, *69*, 134103.

[53] a) Y. Katayama, T. Mizutani, W. Utsumi, O. Shimomura, M. Yamakata, K. Funakoshi, *Nature* **2000**, *403*, 170–173; b) G. Monaco, S. Falconi, W. A. Crichton, M. Mezouar, *Phys. Rev. Lett.* **2003**, *90*, 255701.

[54] L. Pratt, R. E. Richards, *Trans. Faraday Soc.* **1954**, *50*, 670–674.

[55] a) G. Kresse, J. Hafner, *Phys. Rev. B* **1993**, *47*, 558–561; b) G. Kresse, F. J., *Phys. Rev. B* **1996**, *54*, 11169.

[56] a) P. E. Blöchl, *Phys. Rev. B* **1994**, *50*, 17953–17979; b) G. Kresse, D. Joubert, *Phys. Rev. B* **1999**, *59*, 1758.

[57] J. P. Perdew, K. Burke, M. Ernzerhof, *Phys. Rev. Lett.* **1996**, *77*, 3865–3868.

[58] a) J. Heyd, G. E. Scuseria, M. Ernzerhof, *J. Chem. Phys.* **2003**, *118*, 8207−8215; b) J. Heyd, G. E. Scuseria, M. Ernzerhof, *J. Chem. Phys.* **2006**, *124*, 219906; c) J. Paier, M. Marsman, K. Hummer, G. Kresse, I. C. Gerber, J. G. Ángyán, *J. Chem. Phys.* **2006**, *124*, 154709.

[59] a) S. Maintz, V. L. Deringer, A. L. Tchougréeff, R. Dronskowski, *J. Comput. Chem.* **2016**, *37*, 1030–1035; b) R. Nelson, C. Ertural, J. George, V. L. Deringer, G. Hautier, R. Dronskowski, *J. Comput. Chem.* **2020**, *41*, 1931–1940.

[60] a) R. Dronskowski, P. E. Blöchl, *J. Phys. Chem.* **1993**, *97*, 8617–8624; b) V. L. Deringer, A. L. Tchougréeff, R. Dronskowski, *J. Phys. Chem. A* **2011**, *115*, 5461–5466.

[61] A. Stukowski, *Model. Simul. Mater. Sci. Eng.* **2010**, *18*, 015012.

[62] K. Momma, F. Izumi, *J. Appl. Crystallogr.* **2011**, *44*, 1272–1276.

# Supporting Information for

Structure and Bonding in Amorphous Red Phosphorus

Yuxing Zhou,[a] Stephen R. Elliott,[b] and Volker L. Deringer[a,*]

[a] Department of Chemistry, Inorganic Chemistry Laboratory, University of Oxford, Oxford OX1 3QR, United Kingdom

[b] Department of Chemistry, Physical and Theoretical Chemistry Laboratory, University of Oxford, OX1 3QZ, United Kingdom

*E-mail: volker.deringer@chem.ox.ac.uk

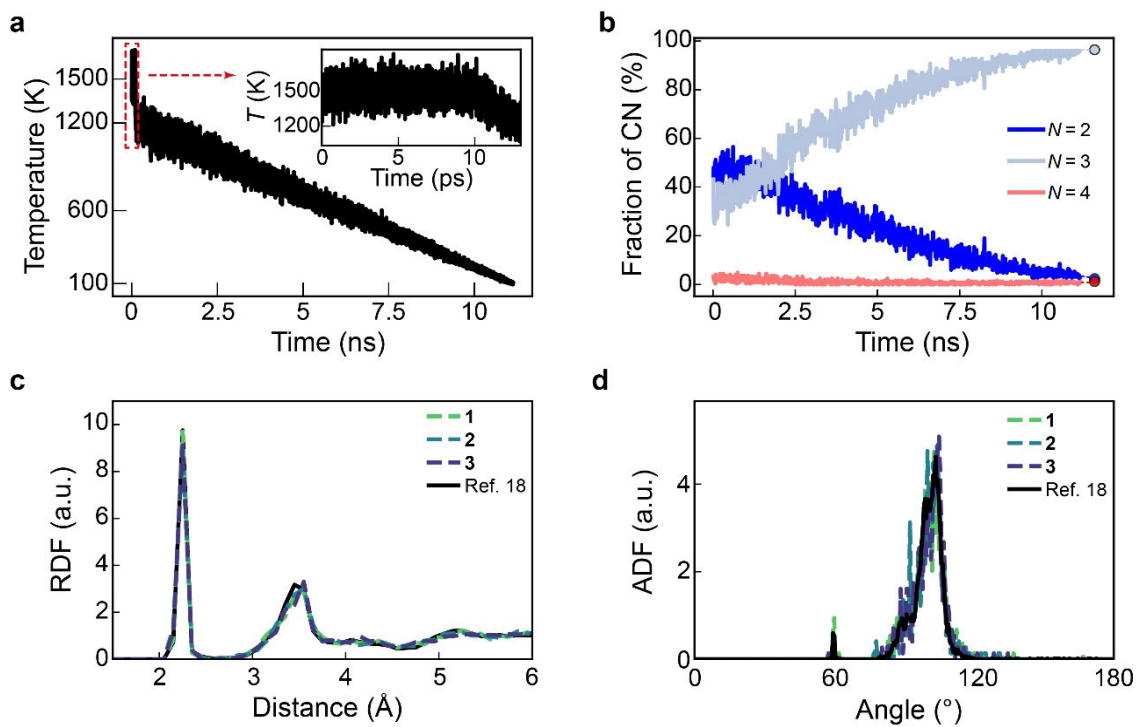

**Figure S1**. Generating a-P structural models with machine-learning-driven simulations. (**a**) The temperature profile of the melt-quench process used to generate the amorphous models in this work (as described in the Methods section): a simulated disordered liquid phase is cooled over about 10 ns, or 10 million simulation steps. (**b**) Counts of coordination numbers during the melt-quench process. In the liquid phase, most atoms are two- ($\approx 49 \pm 4\%$) and three-fold ($\approx 29 \pm 6\%$) connected. During quenching, a network of mostly three-fold-connected atoms emerges, and in the final structures (indicated by markers), almost all of the atoms have $N = 3$ neighbours. (**c**–**d**) Calculated radial distribution function (RDF, panel **c**) and angular distribution function (ADF, panel **d**) for the three pristine models of a-P (**1** to **3**) generated in the present work, compared to those calculated from the larger-scale model in previous work [Y. Zhou, W. Kirkpatrick, V. L. Deringer, *Adv. Mater.* **2022**, *34*, 2107515].

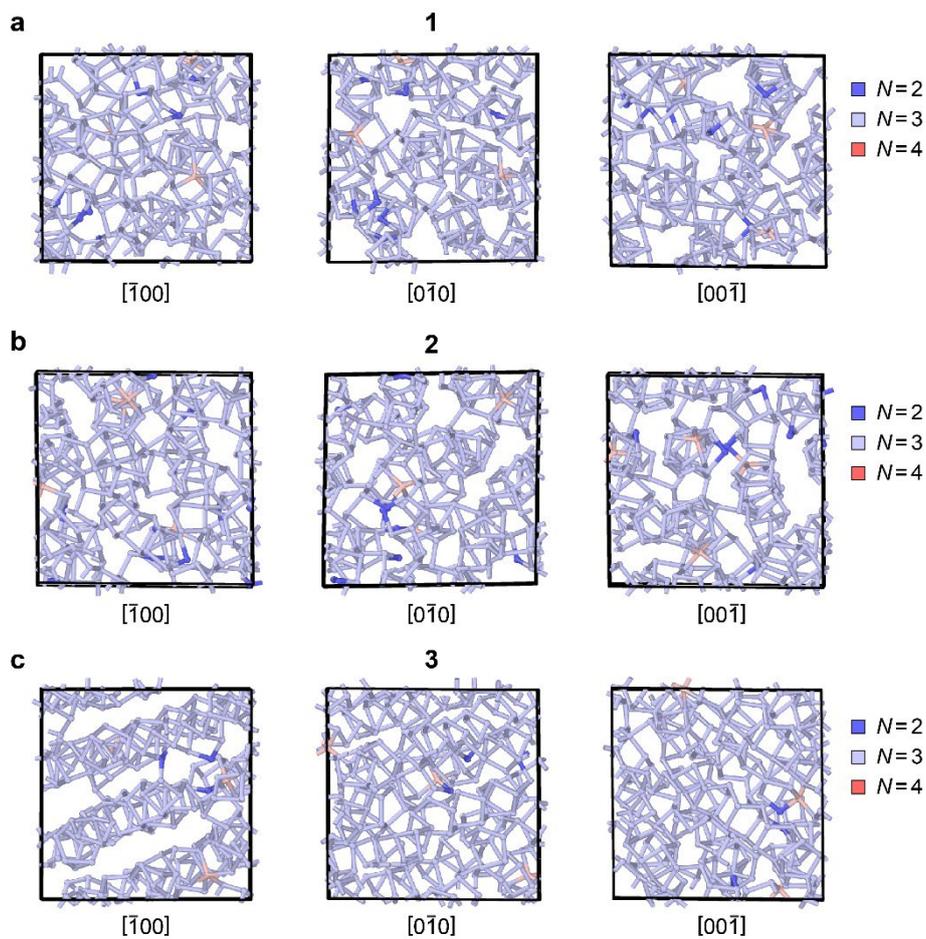

**Figure S2**. (**a–c**) Structures of the three pristine models generated in this work. Views along the [$\bar{1}00$], [$0\bar{1}0$], and [$00\bar{1}$] directions are shown, respectively. Atoms are colour-coded based on the coordination numbers, *viz.* $N = 2$ (blue), $N = 3$ (pale blue), and $N = 4$ (pink), determined by counting atomic neighbours up to a 2.4 Å cut-off.

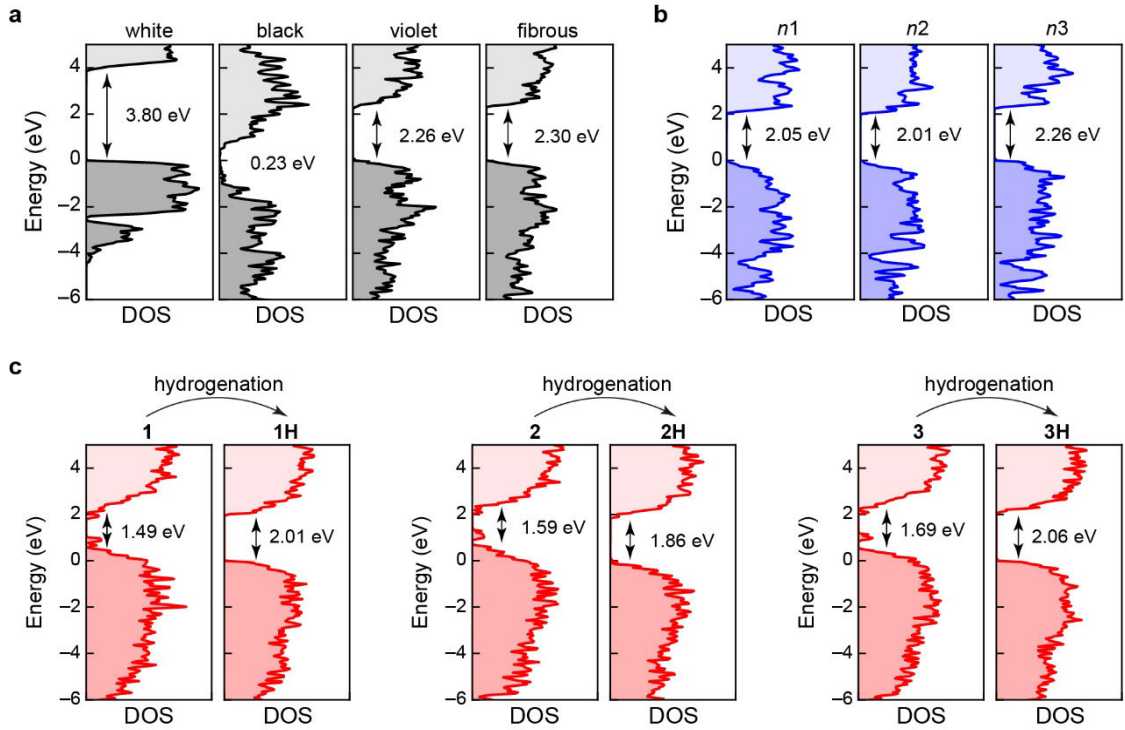

**Figure S3**. Computed electronic densities of states (DOS, at the HSE06 level) for various phosphorus modifications, including (**a**) crystalline phosphorus allotropes (white, black, violet, and fibrous phosphorus), (**b**) three nanorod models (*n*1 to *n*3) following Bachhuber *et al.*, as described in the main text, as well as (**c**) the pristine and hydrogenated a-P models generated in the present work. The band gaps were computed using the difference between the top of the valence band and the bottom of the conduction band: that is, we omitted the mid-gap defect states when determining band gaps for pristine a-P models. An opening of band gaps in the a-P models upon hydrogenation is evident.

S4